\documentclass{article}

\usepackage[utf8]{inputenc}
\usepackage{graphicx}
\usepackage[dvipsnames]{xcolor}
\usepackage{caption}
\usepackage{subcaption}
\usepackage{svg}
\usepackage{amsmath}
\usepackage{mathtools} % extension of amsmath
\usepackage{amssymb} % math symbols
\usepackage{amsthm} % theorem environments
\usepackage{thmtools} % extends amsthm with autoref compat. etc.
\usepackage[subpreambles=false]{standalone}
\usepackage{cleveref}
\usepackage{biblatex}
\graphicspath{ {./res/} }

\usepackage{tikz}
\usepackage{pgfplots}
\usepgfplotslibrary{dateplot}
\pgfplotsset{compat=1.8}

\begin{document}

% -------------------------------------------------------------
% FRONT PAGE
% .............................................................
\begin{titlepage}
\newcommand{\HRule}{\rule{\linewidth}{0.5mm}}
\center
\textsc{\LARGE University of Bergen \\ Department of informatics}\\[1.5cm]
\begin{Huge}
	\bfseries{Latency in Mesh Networks}\\[0.7cm]
\end{Huge}
\large \emph{Author:} Joakim Algrøy\\
\large \emph{Supervisor:} Albin Severinson\\[2cm]
\centerline{\includegraphics[scale=0.15]{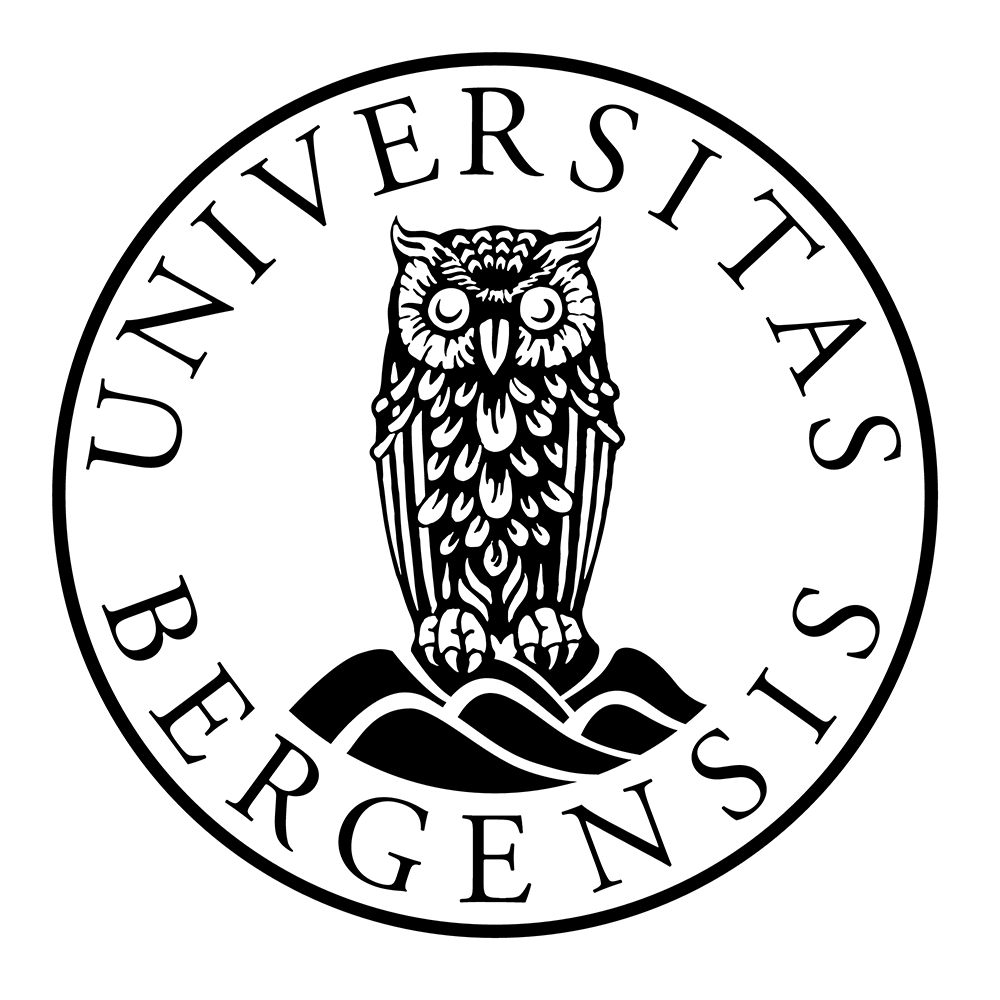}}
{\large December 15, 2020}\\[3cm]
\vfill
\end{titlepage}

% -------------------------------------------------------------
% ABSTRACT
% .............................................................
\pagenumbering{roman}
\begin{abstract} 
\noindent

In this work, we consider latency in wireless networks, motivated by the need for reliable and low-latency wireless communication for applications such as vehicle-to-vehicle communication and industrial automation. The environment that these applications operate in is often highly dynamic, and, as a result, the wireless network technology used needs to be able to adapt to changing conditions, which makes mesh networking a good fit. Hence, in this work, we consider Thread and Bluetooth Mesh, two prominent mesh networking technologies. We measure one-way point-to-point latency using these technologies and show the cumulative probability distribution of the latency for varying payload size. We carry out the measurements under static conditions in an open-air environment. Our results show that Thread achieves lower latency than Bluetooth Mesh, especially for larger payloads. Furthermore, we show that Thread meets some of the latency targets set out for 5G. However, since Thread networks are slow to adapt to network changes, Thread may still not be suitable for application requiring low latency. Bluetooth mesh, on the other hand, does not meet any of the latency targets, although it is close when the payload is small (less than $11$ bytes). Overall, there needs to be further optimization before these technologies can be used for reliable and low-latency communication in dynamic environments.

% In this report, we measure one-way point-to-point latency in wireless networks based on Thread and Bluetooth Mesh, two prominent mesh networking technologies. Our work is motivated by the need for reliable and low-latency wireless communication for applications such as vehicle-to-vehicle communication and industrial automation. We measure latency between two mesh nodes in an open air environment, and show the cumulative probability distribution of these measurements. Our results show lower latency in Thread than in Bluetooth Mesh, especially for larger messages. Thread meets the latency target for some of the discussed applications, however given the slow adaption to network changes in Thread it is not necessarily feasible to use it for those applications. Overall, we show a potential for future improvement in mesh network latency.

\end{abstract}

% -------------------------------------------------------------
% INTRODUCTION
% .............................................................
\section{Introduction}

Reliable and low-latency wireless communication is a critical for many important applications. For example, a vehicle-to-vehicle communication system designed to prevent accidents is dependent on low latency in order to react to dangerous situations in time. Other examples include factory automation, where the latency targets can be less than 1 millisecond (ms) for, e.g., precise robotic motion control, or 10 ms to 50 ms (depending on the application) for cooperative driving \cite{3GPP22804}.
% and 40 ms to 500 ms for standard mobile robot operation and traffic management \cite{3GPP22804}.
% As another example, an Industrial Control System (ICS) might require the ability to shut off or adjust the operation of industrial machines with very little delay. 
However, achieving high reliability and low latency for these applications is challenging. For example, the network topology (including which devices are part of the network) and radio environment may change significantly as vehicles travel or machines are shut on or off.
% Hence, communication systems for these applications need to provide both low latency and resiliency towards changes.

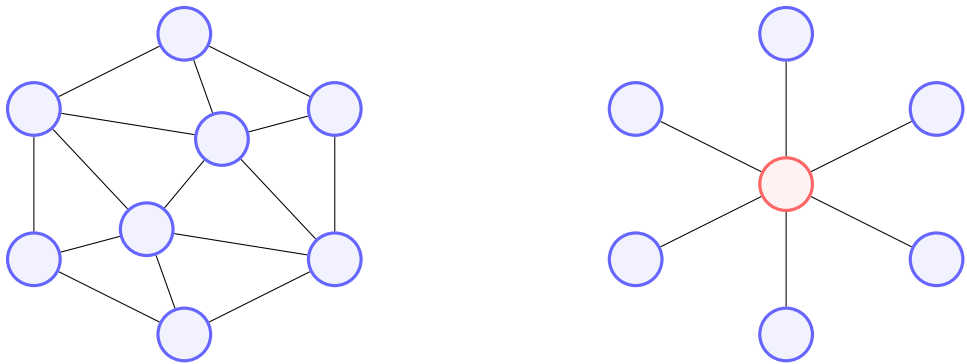
\begin{figure}[h]
  \centering
  \begin{tikzpicture}[
roundnode/.style={circle, draw=blue!60, fill=blue!5, very thick, minimum size=7mm},
routernode/.style={circle, draw=red!60, fill=red!5, very thick, minimum size=7mm},
]
% Mesh Nodes
\foreach \place/\name in {{(0,-1)/ma}, {(2,0)/mb}, {(0.5,1.6)/mc}, 
                          {(0,3)/md}, {(-2,0)/me}, {(-2,2)/mf}, 
                          {(2,2)/mg}, {(-0.5,0.4)/mh}}
    \node[roundnode] (\name) at \place {};
    
%Lines
\draw[-] (ma) -- (mb);
\draw[-] (ma) -- (mh);
\draw[-] (mc) -- (mh);
\draw[-] (me) -- (mh);
\draw[-] (mb) -- (mh);
\draw[-] (mf) -- (mh);
\draw[-] (mb) -- (mc);
\draw[-] (me) -- (mf);
\draw[-] (mf) -- (md);
\draw[-] (ma) -- (me);
\draw[-] (mc) -- (md);
\draw[-] (mc) -- (mf);
\draw[-] (mc) -- (mg);
\draw[-] (mb) -- (mg);
\draw[-] (md) -- (mg);
    
% Star Nodes
\foreach \place/\name in {{(8,-1)/sa}, {(10,0)/sb}, {(6,0)/sc}, 
                          {(8,3)/sd}, {(6,2)/se}, {(10,2)/sf}}
    \node[roundnode] (\name) at \place {};
\node[routernode] (sr) at (8,1) {};

%Lines
\draw[-] (sr) -- (sa);
\draw[-] (sr) -- (sb);
\draw[-] (sr) -- (sc);
\draw[-] (sr) -- (sd);
\draw[-] (sr) -- (se);
\draw[-] (sr) -- (sf);

\end{tikzpicture}
  \caption{Mesh (left) and star (right) network topology. End devices shown in blue and routers in red. Mesh networks do not rely on special devices for routing network traffic. Instead, network nodes collaborate to construct paths through the network.}
  \label{fig:mesh-vs-star}
\end{figure}

A potential solution is provided by mesh networks. In these networks, end devices communicate directly in a peer-to-peer fashion if they are within range, or via intermediary relay nodes (often any node in the network can act as a relay node) otherwise. This is in contrast to so-called \emph{star topology networks} (e.g., WiFi and cellular networks for mobile phones), where communication between end devices has to go through a router, or a hierarchy of routers, whose only role is to relay traffic. We depict the difference in \cref{fig:mesh-vs-star}. Mesh networks are appealing since they require little to no existing infrastructure, which can reduce the cost of deployment, do not depend on routers, which may fail, and are able to adapt to changes by constructing new routes through the network when the environment changes. Furthermore, the direct nature of the communication can also reduce latency.
%Modern wireless networks, such as WiFi and cellular networks, are typically structured using a star topology, where communication between end devices goes through a router or a hierarchy of routers, whose only role is to relay traffic. A mesh network on the other hand is a network with a topology without hierarchy, where nodes in the network do not have a pre-assigned role. In a typical mesh network, any node can communicate directly with any other node given that they are within range of each other. Every node can also act as a router that relays data between nodes that are out of reach of each other.

% Star networks, with a structured hierarchy, are very useful when the primary communication is with geographically distant services, such as most internet communication. However, for communication with nearby entities, mesh networks provide several advantages. Mesh networks require little to no existing infrastructure, which can reduce the cost of deployment. The direct nature of the communication can also reduce latency. Consider, for example, vehicle-to-vehicle communication. Deploying routing devices with coverage along all roads can be very costly, and using cellular towers with large coverage can increase latency due to the amount of traffic each tower must handle. 

Hence, in this report we benchmark existing mesh network technologies to assess how suitable they are for latency-critical applications. In particular, we measure one-way latency between pairs of devices over Thread and Bluetooth mesh networks, which are two of the most prominent mesh network technologies. We find that for 1 byte messages, the median latency is similar for both technologies, however for the 90th and 99th percentile Thread has lower latency than Bluetooth Mesh. For 20 and 100 byte messages, we find that Bluetooth Mesh is significantly impacted by it's fragmentation and re-transmission schemes, and latency is not compatible with latency targets for the discussed applications. For these message sizes Thread is above the 50ms latency target in the 99.9th percentile, but remains within 30ms in the 99th percentile. None if the technologies come close to the 1ms target for precise robotic motion control. 

\subsection{Previous work}

Latency in mesh networks has been measured in several previous reports. In particular, in two reports, published by Silicon Labs and the OpenThread project, respectively.

Silicon Labs published a performance comparison \cite{SiLabsMesh}, where the authors carry out open air tests of Bluetooth Mesh, Thread, and Zigbee in an actively used office building. Note that these technologies use the 2.4GHz frequency band, which is part of the unlicensed spectrum. As a result, a wide variety of different technologies use this spectrum, which will impact the test results. Their test networks ranged in size from 24 to 192 devices, and measured round trip time (as opposed to one-way latency). In the report, the authors conclude that performance is similar across the mesh systems considered for small payloads (e.g. 8 bytes), however as the payload size is increased (e.g., to 50 bytes), Thread has a lower average and less spread in latency than Bluetooth Mesh. The authors attribute this to there being more fragmentation in Bluetooth mesh due to the smaller packet size. Furthermore, the flooding approach of Bluetooth Mesh may cause a higher number of total transmissions in the network, causing delays due to busy channels.

% of Bluetooth Mesh requiring more fragmentation, as well as the flooding approach overcrowding the network with traffic.

The OpenThread project continually measure latency, loss rate, and throughput to evaluate the performance impact of new software versions \cite{OtQualityDash}. The testing method is described in \cite{OtQualityDashPaper}. In particular, the method is based on connecting the devices using coaxial cables, such that the wireless signals are contained within the cables and is isolated from interfering signals from other nearby devices. Testing is done using the nRF52840 Development Kit from Nordic Semiconductors \cite{nrf52840dk}, the same device used in this work. As opposed to the report by Silicon Labs, OpenThread measures one-way latency. This is achieved by using a wired connection between devices to synchronize time. They use a network with 12 devices for measuring loss rate and throughput, but perform latency tests with just 1, 2 and 3 hops (2, 3, and 4 devices, respectively). OpenThread does not run similar tests of other mesh network technologies, so we cannot use their results for direct comparison with Bluetooh Mesh. However, we can compare our OpenThread results with theirs to assess the impact of the differences in testing methods.

\section{An overview of Thread and Bluetooth Mesh}
\label{sec:mesh}

In this work, we consider two prominent wireless mesh network systems; Thread and Bluetooth Mesh. Specifically we consider Google's implementation of Thread, OpenThread \cite{openthread}, which is, for example, used in the Nest line of home automation products, and the implementation of Bluetooth Mesh found in the the Zephyr real-time operating system \cite{zephyrblmesh}. Thread and Bluetooth Mesh differ significantly in their design, both at the physical and protocol layers, which makes comparing their respective performance especially interesting.

Thread's \cite{threadstack} physical and media access control (MAC) layers are based on the IEEE 802.15.4 wireless standard \cite{IEEE802154}. The physical layer uses the 2.4GHz wireless band with a data rate of 250 kbit/s and a frame size of up to 127 bytes. 
% The MAC layer handles acknowledgements and re-transmissions of dropped frames as well as CSMA (Carrier Sense Multiple Access) to avoid collisions. 
Routing in a Thread network is handled by nodes assigned the routing role. These nodes maintain routing tables and forward frames based on the routing information protocol (RIP) algorithm. Every device capable of routing can be assigned the router role when the network deems it necessary, and this happens automatically without user intervention. If a router is no longer needed, it will automatically be demoted to a regular end device. Only keeping the necessary number of routers allows the entire network to remain connected without overcrowding it with relayed traffic. Even though the network reacts to changes dynamically, there can be significant delay before the network reacts, meaning that, e.g., a node leaving the network, can result in long outages.

Bluetooth Mesh \cite{blmeshspec} makes use of the widespread Bluetooth Low Energy (LE) protocol for what is referred to in the specification as the \textit{bearer layer}, which essentially corresponds to the link and physical layers of Thread. Thus, any device designed for Bluetooth LE can also run Bluetooth Mesh with only software modifications. Bluetooth Mesh makes use of the Bluetooth LE advertising channels for communication, which can be set to 1Mbit/s, 500Kbit/s, or 125Kbit/s mode depending on the configuration. The advertising packets have 11 bytes available for the application layer, however 1, 2, or 3 bytes must be used for an opcode, leaving 8 to 10 bytes for application data. Routing is performed by relay nodes in the network, who do not maintain routing tables, but re-transmits any messages that are not intended for them and that meet certain requirements. This approach to routing is referred to as network flooding. Not all nodes need to have relay capability, and among those that do have the relay capability it can be enabled or disabled. As opposed to Thread, enabling and disabling relay capability does not happen automatically in Bluetooth Mesh. Instead this must be manually configured when deploying the network. As shown in \cite{SiLabsBlMesh}, latency in large Bluetooth Mesh networks can be significantly improved by optimizing the number of relay nodes, meaning manual configuration might be necessary to achieve the best possible latency. 

Bluetooth Mesh has several configurable parameters that will have a big impact on latency. To achieve high reliability, Bluetooth Mesh will send each packet multiple times, both when transmitting at the source and when relaying a received packet. Two parameters, network transmit count ($N_\mathsf{TC}$) and Network transmit interval ($N_\mathsf{TI}$) determine how many times a packet is transmitted and the minimum time between the start of each transmission, respectively. In addition to $N_\mathsf{TI}$, a random delay between 0 and 10 ms is added between each packet. While Bluetooth LE v5 supports interleaving packets through \emph{advertising sets}, Bluetooth Mesh remains compatible with Bluetooth LE v4.2 and does not make use of this feature \cite{hernandez2020}. Thus, packets are sent sequentially, leading to a delay of at least $N_\mathsf{TC} \times N_\mathsf{TI}$ between each packet. Given a message requiring \textit{M} packets, a minimum latency when sending directly between two devices (1 hop), can be given by \cref{eq:min-bl-latency}, not including processing delays \cite{hernandez2020}:

\begin{equation}\label{eq:min-bl-latency}
t = T_{trans,0} + \sum_{i=1}^{M-1}\sum_{j=1}^{N_\mathsf{TC}}{(N_\mathsf{TI}+T_{\mathsf{trans},i}+T_{\mathsf{rand},i,j})}
\end{equation}

Where $T_{\mathsf{trans},i}$ is the time it takes to transmit packet $i$, and $T_{\mathsf{rand},i,j}$ is a random value between 0 and 10 ms.

The different approaches to routing will have an impact on latency. In general, we expect routing to be more efficient in Thread compared to Bluetooth Mesh, leading to improved latency. However, Thread may be slow to adapt to changes (e.g. when a router leaves the network) as it needs to reconfigure the routing tables to reflect the new network state. Bluetooth Mesh will not see any such delay, as the nodes do not keep track of routes. Differences in packet sizes will also affect latency. Bluetooth Mesh packets have an application payload of at most 10 bytes, while Thread has a payload of up to 63 bytes. Thus, Bluetooth Mesh will have increased fragmentation compared to Thread, which can lead to higher latency.

% -------------------------------------------------------------
% Main section
% .............................................................

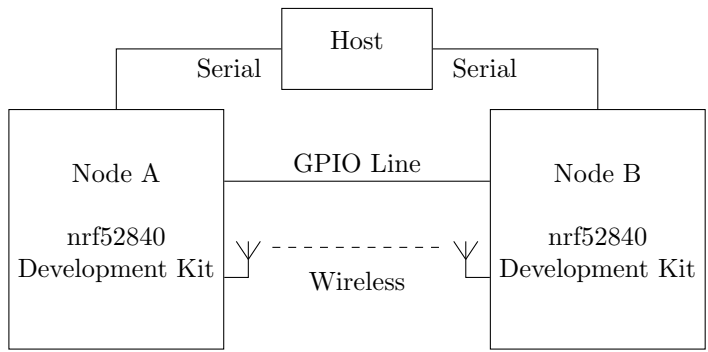
\begin{figure}[t]
  \centering
  \begin{tikzpicture}[
roundnode/.style={circle, draw=blue!60, fill=blue!5, very thick, minimum size=7mm},
routernode/.style={circle, draw=red!60, fill=red!5, very thick, minimum size=7mm},
scale=0.8
]

\node[draw, rectangle, align=center, minimum width=2cm] (h) at (4,3) {\\Host\\};
\node[draw, rectangle, align=center] (a) at (0,0) {\\\\Node A \\ \\nrf52840 \\Development Kit\\\\};
\node[draw, rectangle, align=center] (b) at (8,0) {\\\\Node B \\ \\ nrf52840 \\Development Kit\\\\};

\draw[-] (h.west) -| (a.north) node[below,midway,xshift=15mm]{Serial};
\draw[-] (h.east) -| (b.north) node[below,midway,xshift=-15mm]{Serial};

\draw[-] ([yshift=8mm] a.east) -- ([yshift=8mm] b.west) node[above, midway]{GPIO Line};

\draw[-] ([yshift=-8mm] a.east) -- ([yshift=-8mm, xshift=4mm] a.east) -- ([yshift=-2mm, xshift=4mm] a.east);
\draw[-] ([yshift=-5mm, xshift=4mm] a.east) -- ([yshift=-2mm, xshift=2mm] a.east);
\draw[-] ([yshift=-5mm, xshift=4mm] a.east) -- ([yshift=-2mm, xshift=6mm] a.east);

\draw[-] ([yshift=-8mm] b.west) -- ([yshift=-8mm, xshift=-4mm] b.west) -- ([yshift=-2mm, xshift=-4mm] b.west);
\draw[-] ([yshift=-5mm, xshift=-4mm] b.west) -- ([yshift=-2mm, xshift=-2mm] b.west);
\draw[-] ([yshift=-5mm, xshift=-4mm] b.west) -- ([yshift=-2mm, xshift=-6mm] b.west);

\draw[dashed] ([yshift=-3mm, xshift=8mm] a.east) -- ([yshift=-3mm, xshift=-8mm] b.west) node[below, midway, yshift=-2mm]{Wireless};

\end{tikzpicture}
  \caption{Testing setup with two mesh nodes. A host computer connected via serial is used for test management and to record results.}
  \label{fig:testbed}
\end{figure}

\section{Latency measurements}

We use the nRF52840 development kit \cite{nrf52840dk}, which is based on the nRF52840 system on a chip, for our testing. The nRF52840 supports the physical layer of both Bluetooth Mesh and Thread. There are also implementations of the Thread and Bluetooth Mesh protocol layers available by projects that support this device, namely OpenThread and the Zephyr RTOS, respectively.

Our setup for measuring one-way latency between two devices consists of two nRF52840 devices in close proximity, connected via a GPIO line. The system is depicted in \cref{fig:testbed}. The following procedure is used for each individual latency measurement:
\begin{enumerate}
    \item The transmitting device sends a signal, via the wired GPIO line to the receiving device. Immediately afterwards the transmitting device sends a mesh network message.
    \item The GPIO signal triggers a hardware interrupt on the receiving device, which immediately records a timestamp indicating the time of transmission for the current packet.
    \item The receiving device records a new timestamp once it has received the wireless mesh message. The difference between these two timestamps is recorded as the one-way latency of that message.
    \item We wait 500 ms, and then return to step 1.
\end{enumerate}  

This testing process is performed for both mesh technologies, using messages of 1, 20, and 100 bytes.

\begin{figure}[t]
    \centering
    \begin{tikzpicture}
\begin{axis}[
height=6cm,
legend cell align={left},
legend style={at={(0,1.3)}, anchor=north west},
log basis x={10},
log basis y={10},
tick align=outside,
tick pos=left,
width=6cm,
grid style={gray,opacity=0.5,dotted},
xlabel={Time $t$ [ms]},
xmajorgrids,
xmin=0, xmax=50,
xminorgrids,
xmode=linear,
ylabel={$\Pr(\text{latency} > t)$},
ymajorgrids,
ymin=0.001, ymax=1,
yminorgrids,
ymode=log,
table/col sep=comma,
cycle list name=exotic,
legend style={cells={align=left}},
]
  % OpenThread 1 Byte
  \addplot+ [PineGreen, mark=*, mark size=2, mark repeat=50, mark options={fill=white}]
  table[x expr=\thisrowno{0}/1000, y expr=1-\thisrowno{1}]{res/openthread/openthread_1B_cdf_100k.csv};
  \addlegendentry{1 byte OpenThread}
  % Bluetooth Mesh 1 Byte
  \addplot+ [Orange, mark=square*, mark size=2, mark repeat=25, mark options={fill=white}]
  table[x expr=\thisrowno{0}/1000, y expr=1-\thisrowno{1}]{res/bluetooth-mesh/bluetooth_1B_cdf_100k.csv};
  \addlegendentry{1 byte Bluetooth Mesh}

\end{axis}
%\node at (2.2,4.9) {1 byte messages};
\end{tikzpicture}
\begin{tikzpicture}
\begin{axis}[
height=6cm,
legend cell align={left},
legend style={at={(0,1.3)}, anchor=north west},
log basis x={10},
log basis y={10},
tick align=outside,
tick pos=left,
width=6cm,
grid style={gray,opacity=0.5,dotted},
xlabel={Time $t$ [ms]},
xmajorgrids,
xmin=0, xmax=300,
xminorgrids,
xmode=linear,
ymajorgrids,
ymin=0.001, ymax=1,
yminorgrids,
ymode=log,
ymajorticks=false,
yminorticks=false,
%ytick pos=right,
table/col sep=comma,
cycle list name=exotic,
legend style={cells={align=left}},
]
  % OpenThread 20 Bytes
  \addplot+ [NavyBlue, mark=*, mark size=2, mark repeat=50, mark options={fill=white}]
  table[x expr=\thisrowno{0}/1000, y expr=1-\thisrowno{1}]{res/openthread/openthread_20B_cdf_100k.csv};
  \addlegendentry{20 byte OpenThread}
  % Bluetooth Mesh 20 Bytes
  \addplot+ [Sepia, mark=square*, mark size=2, mark repeat=5, mark options={fill=white}]
  table[x expr=\thisrowno{0}/1000, y expr=1-\thisrowno{1}]{res/bluetooth-mesh/bluetooth_20B_cdf.csv};
  \addlegendentry{20 byte Bluetooth Mesh}
\end{axis}
%\node at (2.2,4.9) {20 byte messages};
\end{tikzpicture}
    \caption{Results of latency tests with 1 and 20 byte messages. 100 000 packets with 250ms interval between receive and next transmission.}
    \label{fig:1_20_cdf_comparison}
\end{figure}
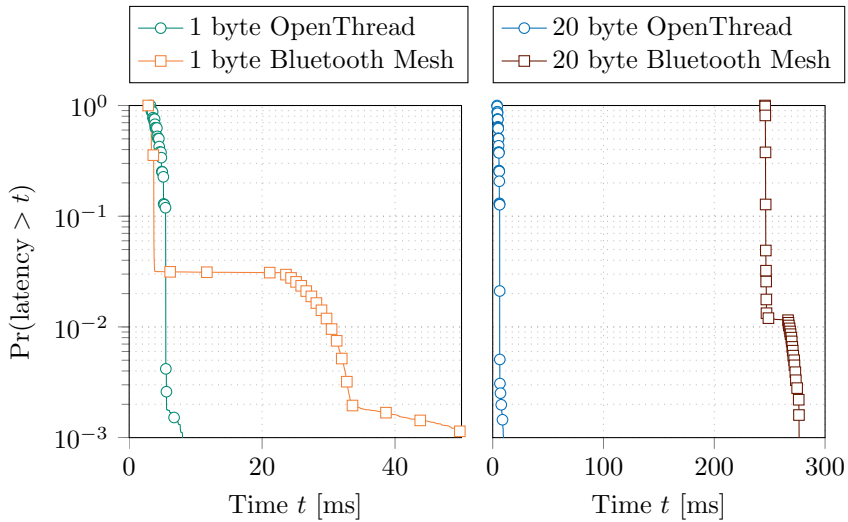

\begin{figure}[t]
    \centering
    \begin{tikzpicture}
\begin{axis}[
height=6cm,
legend cell align={left},
legend style={at={(0,1.2)}, anchor=north west},
log basis x={10},
log basis y={10},
tick align=outside,
tick pos=left,
width=6cm,
grid style={gray,opacity=0.5,dotted},
xlabel={Time $t$ [ms]},
xmajorgrids,
xmin=0, xmax=60,
xminorgrids,
% xmode=log,
ylabel={$\Pr(\text{latency} > t)$},
ymajorgrids,
ymin=0.001, ymax=1,
yminorgrids,
ymode=log,
table/col sep=comma,
cycle list name=exotic,
legend style={cells={align=left}},
]
  % OpenThread 100 Bytes
  \addplot+ [orange, mark=*, mark size=2, mark repeat=50, mark options={fill=white}]
  table[x expr=\thisrowno{0}/1000, y expr=1-\thisrowno{1}]{res/openthread/openthread_100B_cdf.csv};
  \addlegendentry{OpenThread}
\end{axis}
\end{tikzpicture}
\begin{tikzpicture}
\begin{axis}[
height=6cm,
legend cell align={left},
legend style={at={(0,1.2)}, anchor=north west},
log basis x={10},
log basis y={10},
tick align=outside,
tick pos=left,
width=6cm,
grid style={gray,opacity=0.5,dotted},
xlabel={Time $t$ [ms]},
xmajorgrids,
xmin=900, xmax=1100,
xminorgrids,
% xmode=log,
%ylabel={$\Pr(\text{latency} > t)$},
ymajorgrids,
ymin=0.001, ymax=1,
yminorgrids,
ymode=log,
%ytick pos=right,
ymajorticks=false,
yminorticks=false,
table/col sep=comma,
cycle list name=exotic,
legend style={cells={align=left}},
]
  % Bluetooth Mesh 100 Bytes
  \addplot+ [mark=square*, mark size=2, mark repeat=15, mark options={fill=white}]
  table[x expr=\thisrowno{0}/1000, y expr=1-\thisrowno{1}]{res/bluetooth-mesh/bluetooth_100B_cdf.csv};
  \addlegendentry{Bluetooth Mesh}
\end{axis}
\end{tikzpicture}
    \caption{Results of latency tests with 100 byte messages. 10 000 packets with 500ms interval. Note the different scale on the time axis.}
    \label{fig:100B_cdf_comparison}
\end{figure}
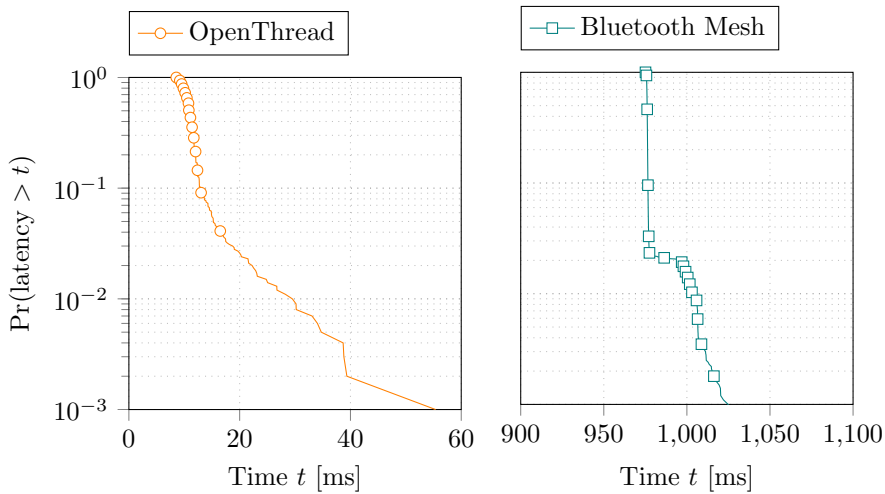

\subsection{Latency distribution}

As seen in \cref{fig:1_20_cdf_comparison}, for 1 byte messages, median latency is similar for Bluetooth Mesh and Thread. However, Bluetooth Mesh has a worse tail latency than Thread. For both technologies, we see that the latency target of 1 ms for precise robotic motion control is not in reach, even for this minimal message size. However, OpenThread remains within 10ms with $> 99.9\% $ probability, and Bluetooth Mesh remains within 50ms with close to $99.9\% $ probability, which was mentioned in the introduction as the lower and upper end of a range of latency targets for cooperative driving.

Once we increase the message size to 20 bytes for Bluetooth Mesh, we see the results of the transmission delays between packets discussed in the introduction. A 20 byte message with a 3-byte opcode (3 bytes needed for vendor-specific opcodes, see \cref{sec:mesh}), will need 3 packets to be transmitted with Bluetooth Mesh. We have set parameters $N_\mathsf{TC} = 3$ and $N_\mathsf{TI} = 20$, giving an absolute minimum latency of 120 ms by \cref{eq:min-bl-latency}, assuming all random delays are zero. However, looking at the source code of the Zephyr Project's implementation of Bluetooth Mesh, we see that the actual time used to send a packet is $30 + (N_\mathsf{TI} + 10)\cdot N_\mathsf{TC}$ \cite{ZephyrSource}, indicating that 240 ms is the minimum expected latency, which is also clear from the 20 byte plot in \cref{fig:1_20_cdf_comparison}. OpenThread maintains a low latency, and is still within 10ms with $> 99.9\% $ probability for 20 byte messages.

\subsection{Latency time series}

\begin{figure}[t]
  \centering
  \begin{subfigure}[b]{0.49\linewidth}
    \centering
    \includegraphics[width=\textwidth]{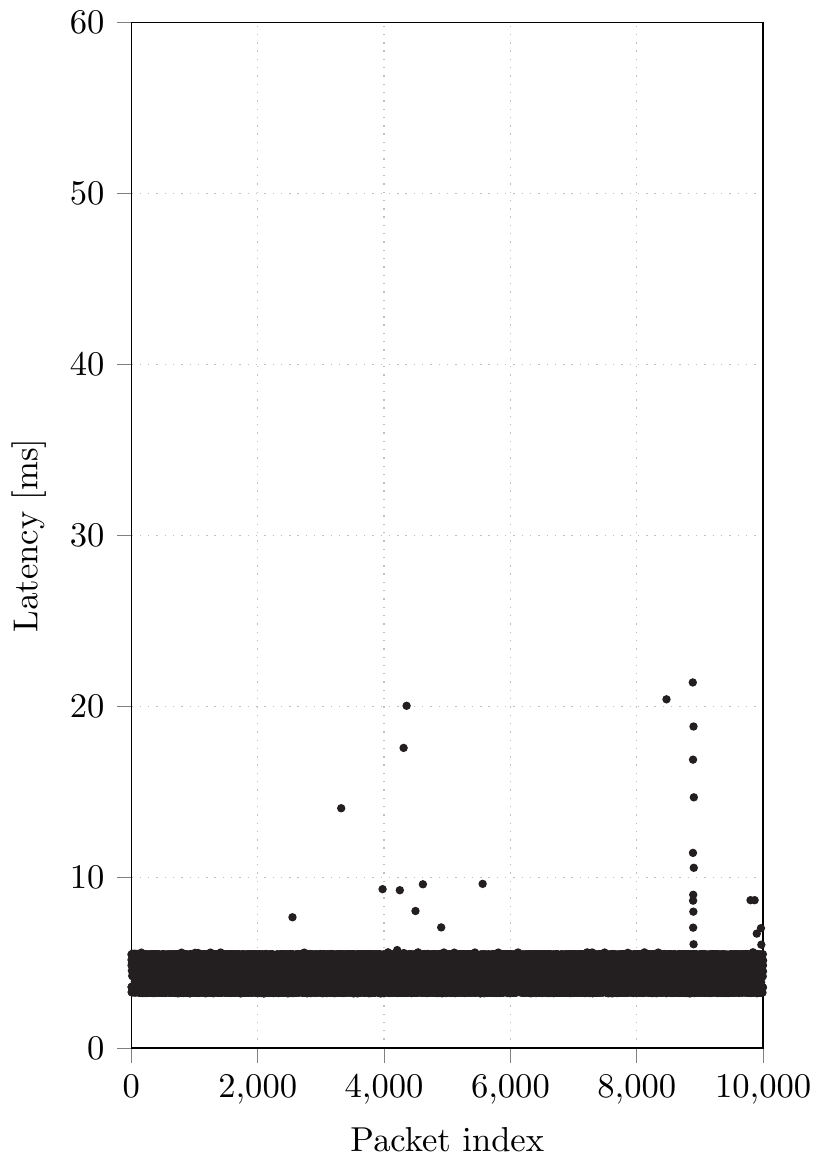}
    \caption{OpenThread}
    \label{fig:openthread_time_series}
  \end{subfigure}
  \hfill
  \begin{subfigure}[b]{0.49\linewidth}
    \centering
    \includegraphics[width=\textwidth]{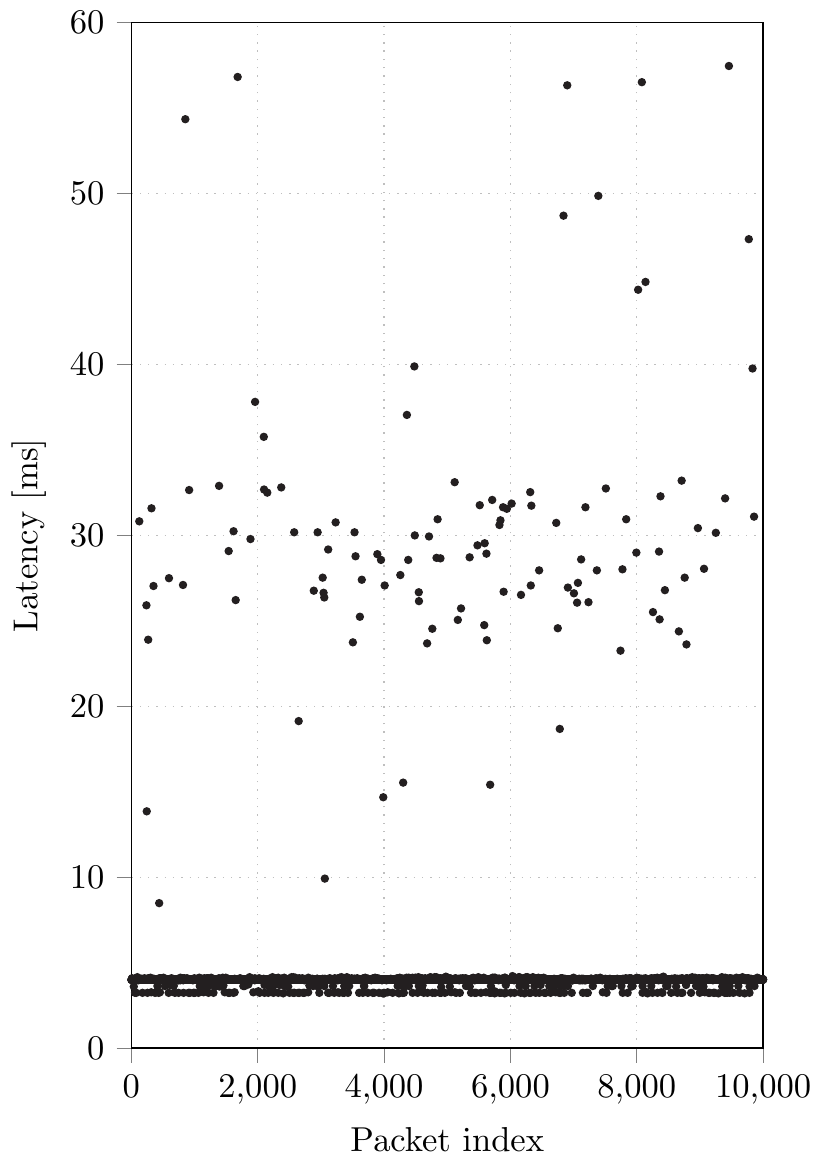}
    \caption{Bluetooth Mesh}
    \label{fig:bluetooth_time_series}
  \end{subfigure}
  \caption{Latency of 1 byte packets in order.}
  \label{fig:time_series}
\end{figure}

In \cref{fig:time_series} we plot latency of 1 byte messages in the measured order for OpenThread and Bluetooth Mesh. In both cases, we see a clear range of high density where the most common latency results lie. We also see a number of measurements above the dense range, which seem to be more uniformly distributed in Bluetooth Mesh, whereas in OpenThread they are more sparse and appear in bursts. These bursts may correspond to changes in interference from the environment.

\section{Conclusions}

Low-latency wireless mesh networks can be an important technology for enabling applications such as cooperative autonomous driving and factory automation since the network is able to automatically adapt to changing conditions. However, meeting the latency targets set out for these applications is challenging. In our results, Thread maintains a one-way single-hop latency below 10ms for messages with a payload of up to 20 bytes with $> 99.9\% $ probability, which is within the the 10-50ms latency targets for cooperative driving suggested in \cite{3GPP22804}. However, given that Thread is slow in adapting to changes in network topology (which occur, e.g., when a vehicle moves out of range), it may be unfeasible to use for applications such as such as cooperative driving. Bluetooth Mesh does not have this limitation, however we show that latency is significantly higher than for Thread. For example, for messages with a 1 byte payload, the $99.9$-th percentile latency is about 50ms. For messages with a payload larger than 11 bytes, the latency for Bluetooth mesh is more than an order of magnitude larger than that of Thread due to latency caused by fragmentation and re-transmission. For example, the minimum latency observed for Bluetooth mesh and messages with a 20 and 100 byte payload is 240ms and 950ms, respectively. 

Our results indicate that Bluetooth Mesh is only suitable for applications requiring very small message sizes (less than 11 bytes), or without stringent latency requirements. Thread, on the other hand, is suitable for applications requiring a $99.9$-th percentile latency of up 10 ms or more when the payload size is at most 20 bytes, or a $99.9$-th percentile latency of about 50 ms or more when the payload is at most 100 bytes. However, Thread is only suitable for applications operating in a static environment, due to the fact that it is slow to adapt to changing conditions. Hence, neither technology is able to fully meet the requirements set out for, e.g., cooperative autonomous driving.

% Overall, we see that while Thread performs well in our static testing environment, we are aware of it's limitations in a more dynamic environment. Bluetooth Mesh should not have those same limitations, however we see that Bluetooth Mesh exceeds our discussed latency targets with an order of magnitude for larger packet sizes. A mesh network achieving the latency of Thread, but using a more dynamic routing protocol could be very well suited for applications such as vehicle-to-vehicle communication or certain applications in factory automation.

% We measure a latency of $> 240$ms for all 20 byte messages, and a latency of $> 950$ms for all 100 byte messages. This illustrates that Bluetooth Mesh is best suited for applications requiring very small message sizes, or very relaxed latency requirements.

% Once messages are longer than what will fit in a single packet (at most 11 bytes), we find a very significant increase in latency due to the re-transmission scheme in Bluetooth Mesh. We measure a latency of $> 240$ms for all 20 byte messages, and a latency of $> 950$ms for all 100 byte messages. This illustrates that Bluetooth Mesh is best suited for applications requiring very small message sizes, or very relaxed latency requirements.

% -------------------------------------------------------------
% REFERENCES
% .............................................................
\clearpage
\printbibliography

\end{document}